\documentclass[superscriptaddress,
twocolumn,showpacs,preprintnumbers,amsmath,amssymb]{revtex4}
\usepackage{graphicx}
\usepackage{dcolumn}
\usepackage{bm}
\usepackage{latexsym}
\begin{document}
\title{Interacting RNA polymerase motors on DNA track: \\
effects of traffic congestion and intrinsic noise on RNA synthesis}
\author{Tripti Tripathi}
\affiliation{Physics Department, Indian Institute of Technology, Kanpur 208016, India}
\author{Debashish Chowdhury}
\affiliation{Physics Department, Indian Institute of Technology, Kanpur 208016, India, and\\
Max-Planck Institute for Physics of Complex Systems, 01187 Dresden, Germany}
\email{debch@iitk.ac.in}
\date{\today}%
\begin{abstract} 
RNA polymerase (RNAP) is an enzyme that synthesizes a messenger RNA (mRNA) 
strand which is complementary to a single-stranded DNA template. From the 
perspective of physicists, an RNAP is a molecular motor that utilizes 
chemical energy input to move along the track formed by a DNA. In many  
circumstances, which are described in this paper, a large number of RNAPs 
move simultaneously along the same track; we refer to such collective 
movements of the RNAPs as RNAP traffic. Here we develop a theoretical 
model for RNAP traffic by incorporating the steric interactions between 
RNAPs as well as the mechano-chemical cycle of individual RNAPs during 
the elongation of the mRNA. By a combination of analytical and numerical 
techniques, we calculate the rates of mRNA synthesis and the average 
density profile of the RNAPs on the DNA track. We also introduce, and 
compute, two new measures of fluctuations in the synthesis of RNA. 
Analyzing these fluctuations, we show how the level of {\it intrinsic 
noise} in mRNA synthesis depends on the concentrations of the RNAPs as 
well as on those of some of the reactants and the products of the 
enzymatic reactions catalyzed by RNAP. We suggest appropriate experimental 
systems and techniques for testing our theoretical predictions. 
\end{abstract}
\maketitle
\section{Introduction} 

Molecular motors \cite{schliwa,hackney,fisher} are either proteins or 
macromolecular complexes that utilize some form of input energy (often 
chemical energy) to perform mechanical work. In many circumstances, 
molecular motors move collectively on a single track in a manner that 
has strong resemblance with vehicular traffic \cite{polrev,physica}. 
In recent years some minimal models of molecular motor traffic have 
been developed to study their generic features 
\cite{lipo,frey,santen,popkov1}. 
More detailed models for specific motor traffic systems have also been 
proposed by capturing the stochastic mechano-chemistry of individual 
motors as well as their steric interactions within the same model to 
investigate the interplay of individual and collective dynamics of the 
motors \cite{basuchow,nosc,greulich}. In this paper we develop such a 
model for a specific class of motors for which no attempt has been made 
in the past to capture their steric interactions during traffic-like 
collective movements on a single track.

According to the central dogma of molecular biology, the genetic 
message stored in the DNA is first {\it transcribed} into messenger 
RNA (mRNA) from which it is then {\it translated} into proteins.
Polymerization of a mRNA from the corresponding single-stranded DNA 
(ssDNA) template is carried out by a motor called RNA polymerase (RNAP) 
\cite{wangrev,landick98a,korzheva03}. In contrast, synthesis of a 
protein from the corresponding mRNA template is mediated by another 
motor, called ribosome, which translocates along the mRNA strand. 
The steric interactions between the neighbouring ribosomes, which 
simultaneously translocate along the same mRNA, were taken into 
account in most of the theoretical models of translation developed 
since the late sixties 
\cite{macdonald68,macdonald69,lakatos03,shaw03,shaw04a,shaw04b,chou03,chou04,schon04,schon05,dong,basuchow}.
Surprisingly, in spite of the close similarities between the 
template-dictated and motor-driven polymerization of macromolecules in 
transcription and translation, no attempt has been made in the past  
to incorporate interactions of RNAPs in the theoretical description of 
transcription. Instead, to our knowledge, all the models of transcription 
reported so far  
\cite{julicherrnap98,osterrnap98,sousa96,sousa97,sousa06,mdwang04,mdwang07,nudler05,tadigotla,peskinrnap06,woo06,liverpool} 
capture only the stochatic mechano-chemistry of the individual RNAP 
motors. Cooperation and collisions between RNAP motors is known to 
have non-trivial effects on the rate of transcription 
\cite{nudler03a,nudler03b,crampton06,sneppen05}. Moreover, the 
possibility of the formation of queues in RNAP traffic has also been 
explored \cite{bremer95}. In fact, if the gene is relatively short, 
a sufficiently long queue of RNAPs on the ssDNA template can reduce the 
accessibility of the promoter sequence thereby lowering the rate of 
further initiation of transcription.

The main aim of this paper is to develop a model of RNAP traffic that 
incorporates steric interactions between RNAP motors which move along 
the same DNA track. In this model, we incorporate the most essential 
features of the multi-step mechano-chemical pathway of the individual 
RNAP motors by a scheme which was used earlier in  Wang et al.'s 
\cite{osterrnap98} model for single RNAP. The steric interaction 
between the RNAPs is assumed to be hard-core repulsion. The effects 
of these interactions of RNAPs is captured in our model of mRNA 
synthesis  in the same manner in which the steric interactions of 
ribosomes was captured in a recent model \cite{basuchow} of protein 
synthesis. 

In the spirit of traffic science \cite{css}, we define the {\it flux} 
to be the average number of motors crossing a site per unit time. 
Thus, flux is expressed in the units ``number per second''. We define 
the number density to be the average number of RNAPs attached to unit 
length of the DNA template. Using the terminology of traffic science, 
we refer to the relation between the flux and the number density of 
the RNAPs as the {\it fundamental diagram}. We calculate the flux and 
investigate its dependence on the number density of RNAP on the DNA 
template as well as on some other experimentally accessible parameters 
of the model. Since average speed of a RNAP is also a measure of the 
average rate of mRNA elongation and the flux gives the total rate of 
mRNA synthesis from a DNA template, our calculations predict the 
effects of RNAP traffic congestion on the rate of synthesis of mRNA.

The steps of the mechano-chemical cycle of a RNAP are intrinsically 
stochastic and give rise to fluctuations in the rates of synthesis 
of mRNA. We introduce quantitative measures of these fluctuations 
by drawing analogy with some further concepts from traffic science 
\cite{css}. We define the {\it run time} $T$ of a RNAP to be the 
actual time it takes to travel from the start site to the stop site 
on the DNA template (i.e., the time taken to synthesize an mRNA 
transcript). Similarly, we define the {\it time-headway} $\tau$ to 
be the time interval between the departures of two successive RNAPs 
from the stop site on the DNA template (i.e., the time interval 
between the completion of the synthesis of successive mRNA transcripts). 
Using the stochastic model which we develope here for RNAP traffic, 
we also compute the distributions $\tilde{P}_{T}$ and ${\cal P}_{\tau}$ 
of run-times and time-headways respectively.

In recent years, stochasticity in gene expression has been probed by 
novel experimental techniques and the results have inspired several 
theoretical models at different levels of complexity \cite{kaern05}. 
The cell-to-cell variations in the levels of expression of the same 
gene can arise from inherently intrinsic fluctuations in 
transcription and translation or from extrinsic causes 
\cite{paulsson05}. Since proteins are the final products of gene 
expression, normally, fluctuations in the concentration of proteins 
are taken as a measure of the noise in gene expression. However, the 
most direct way to measure transcriptional noise would be to monitor 
the fluctuations in the synthesis of mRNA transcripts 
\cite{golding05,raj,chubb06,darzacq07a}. 
Therefore, instead of modeling cell-to-cell variations in the 
transcription of a specific gene, in this paper we study the  
RNAP-to-RNAP fluctuations in the synthesis of mRNA from a  
single DNA template. The width of the distributions $\tilde{P}_{T}$ 
and ${\cal P}_{\tau}$ provide measures of the contributions to  
transcriptional noise from the intrinsic fluctuations in the steps 
of the mechano-chemical cycle of RNAPs on the same DNA template.

The paper is organized as follows. In section \ref{sec-bioold} we 
summarize the essential mechano-chemical processes involved in 
transcription. In the same section we also present a brief review 
of some of the relevant earlier models. Our stochastic model is 
developed in section \ref{sec-model}. Our theoretical predictions 
on flux and average density profiles, which follow from this model 
under periodic and open boundary conditions, are discussed in the 
sections \ref{sec-pbc} and \ref{sec-obc}, respectively. Our results 
on fluctuations and transcriptional noise are presented in section 
\ref{sec-fluc}. The experimental implications of our theoretical 
predictions are discussed in section \ref{sec-expt}.
Finally, in section \ref{sec-conclude} we summarize our main 
theoretical predictions.

\section{Brief review of phenomenology and earlier models} 
\label{sec-bioold}

\subsection{Essential chemo-mechanical processes} 

DNA and RNA are linear polymers whose monomeric subunits are called 
nucleotides. Transcription, i.e., the process of synthesis of mRNA from 
the corresponding ssDNA template, can be broadly divided into three 
stages, namely, {\it initiation}, {\it elongation} and {\it termination}. 
In the initiation stage, an RNAP recognizes the so-called ``promoter 
sequence'' on the DNA and locally unzips the two DNA strands creating 
a ``bubble'' whereby a ssDNA template is exposed to it. However, in 
this paper we are interested mainly in the elongation of the mRNA 
transcript. 

During elongation \cite{uptain}, each successful addition of a 
nucleotide to the elongating mRNA leads to a forward stepping of the 
RNAP. The RNAP, together with the DNA bubble and the growing RNA 
transcript, forms a ``transcription elongation complex'' (TEC). The 
essential components of each of the TECs are shown explicitly in the 
schematic depiction of RNAP traffic in fig.\ref{fig-rnaptraf}(a). 
As reported in the literature \cite{osterrnap98}, the typical size of 
a transcription bubble is about $15$ nucleotides (i.e., about $5$ 
nm) whereas a single RNAP covers a DNA segment that can be as long as 
$35$ nucleotides (i.e., about $12$ nm). The non-template DNA strand 
remains in single-stranded conformation in the bubble region while 
a $8$-$10$ nucleotide-long DNA-RNA hetero-duplex is formed by a 
part of the template DNA strand and the growing end of the RNA (see 
fig.\ref{fig-rnaptraf}(a)).

Each mechano-chemical cycle of the RNAP during the elongation stage 
\cite{wangrev,cramer02a,cramer05} consists of several steps; the major 
steps being (i) Nucleoside triphosphate (NTP) binding to the active 
site of the RNAP when the active site is located at the growing tip of 
the mRNA transcript, (ii) NTP hydrolysis, (iii) release of pyrophosphate 
($PP_{i}$), one of the products of hydrolysis, and (iv) accompanying 
forward stepping of the RNAP \cite{wangrev}. This simplified scenario, 
which is adequete for our purpose here, is shown symbolically in 
equation (\ref{eq-cycle}):
\begin{widetext}
\begin{eqnarray} 
TEC_{n} + NTP \leftrightharpoons  TEC_{n} \bullet NTP \leftrightharpoons TEC_{n+1} \bullet PP_{i} \leftrightharpoons TEC_{n+1}
\label{eq-cycle}
\end{eqnarray}
\end{widetext}
The elongation process ends when the TEC encounters the corresponding 
``termination sequence'' and the nascent mRNA is released by the RNAP.

\subsection{Brief review of the earlier models} 

A stochastic chemical kinetic model was developed by J\"ulicher and 
Bruinsma \cite{julicherrnap98} to describe not only the polymerization of 
mRNA by a RNAP, but also to account for the effects of elastic strain 
in the motor. Almost simultaneously, Wang et al.\cite{osterrnap98} developed 
a model that incorporated the multi-step chemical kinetics of the 
transcription elongation process. Extending Von Hippel's 
\cite{hippel92,vonhippel98,hippel05} 
pioneering works on sequence-dependent thermodynamic analysis of 
transcription, Wang and collaborators \cite{mdwang04,wangrev} have developed 
a sequence-dependent kinetic model in terms of a transcription-energy 
landscape. This model has been extended further by Tadigotla et al. 
\cite{tadigotla} by incorporating the kinetic barriers erected by the 
folding of the mRNA transcript. Very recently, Bai et al.\cite{mdwang07} 
have demonstrated the predictive power of their theoretical model carrying 
out experiment and data analysis in two stages: in the first they 
estimated the model parameters from experimental exploration of the 
response to chemical perturbations, and, then, in the second stage 
using these parameters they predicted the responses to mechanical 
perturbations. But, as stated in the introduction, none of these models 
incorporate steric interactions between the RNAPs.

To our knowledge, the first model of molecular motor traffic was 
developed almost forty years ago by MacDonald, Gibbs and collaborators 
\cite{macdonald68,macdonald69} in the context of ribosome traffic. 
In the pioneering works \cite{macdonald68,macdonald69}, as well as 
in most of the extensions in recent years 
\cite{lakatos03,shaw03,shaw04a,shaw04b,chou03,chou04,schon04,schon05,dong},
the details of molecular composition and achitecture as well as the 
mechano-chemical cycles of the ribosomes were not taken into account. 
Instead, each ribosome was modelled as a hard rod; in the special limit 
where the size of the rod coincides with the lattice constant, this model 
reduces to the totally asymmetric simple exclusion process (TASEP) which 
is the simplest model of interacting self-propelled particles. Very 
recently, a more realistic model \cite{basuchow} of ribosome traffic has 
been developed by incorporating the essenial steps in the mechano-chemical 
cycle of a ribosome during the elongation of the protein. Traffic of some 
other families of motors have also been modelled recently in the same 
spirit, i.e., by incorporating both the intra-motor mechano-chemistry and 
inter-motor steric interactions \cite{nosc}.

\section{Model}
\label{sec-model}

\begin{figure}[h]
\begin{center}
(a)\\[2.25cm]
\includegraphics[angle=-90,width=0.85\columnwidth]{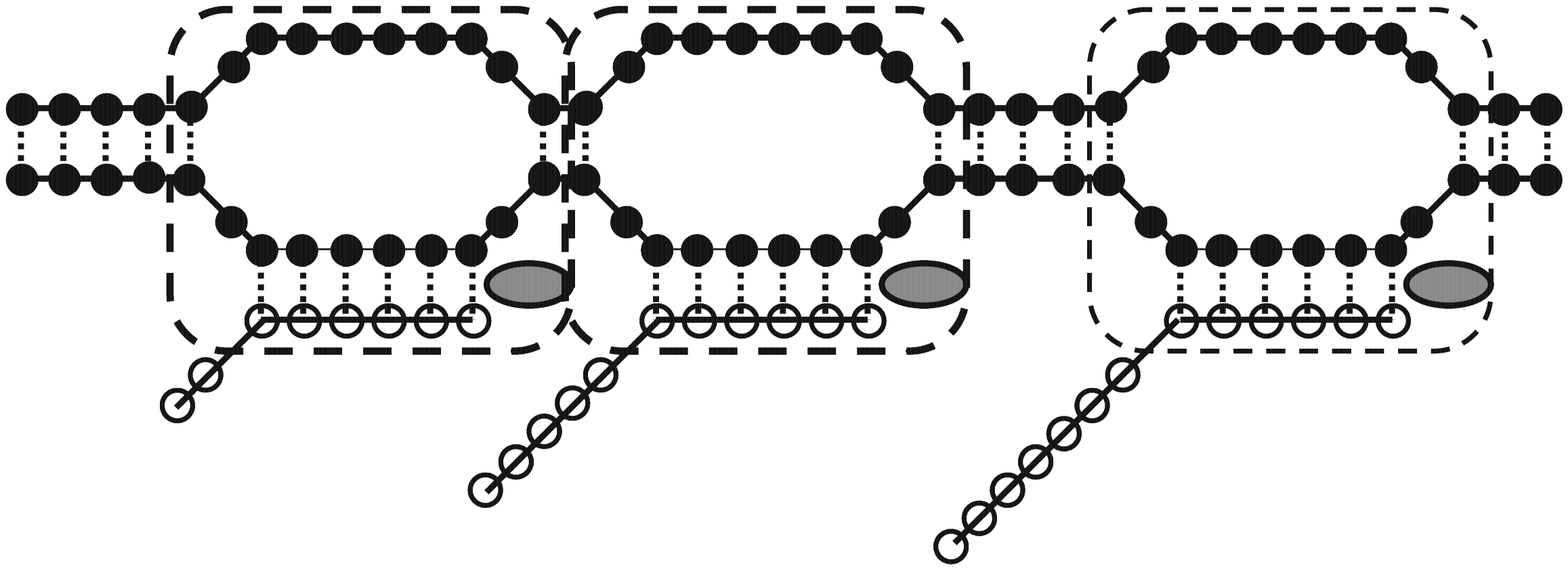}\\[0.5cm]
(b)\\[0.5cm]
\includegraphics[angle=-90,width=0.85\columnwidth]{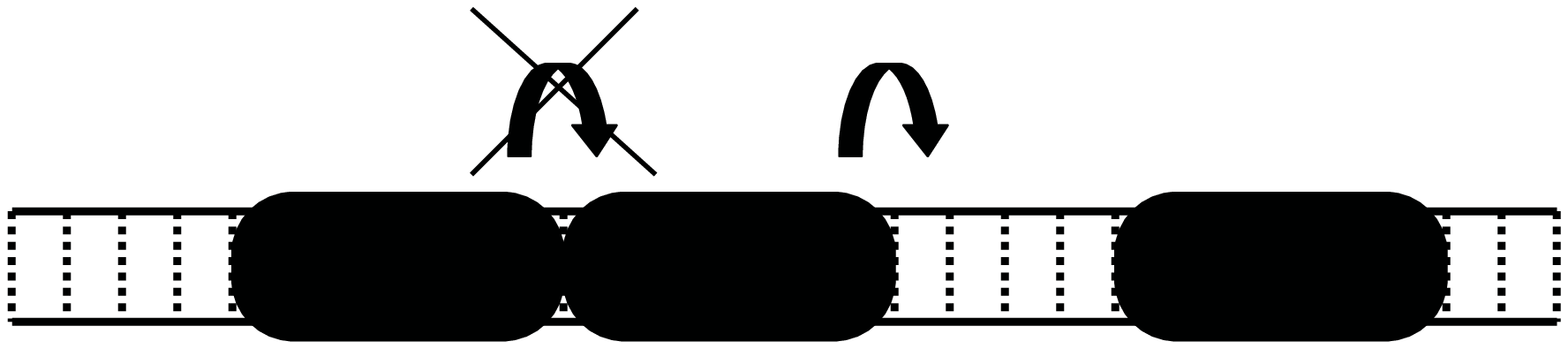}
\end{center}
\caption{(a) A schematic representation of RNAP traffic where the three 
dashed squares represent three TECs. The solid lines connecting filled 
circles represent the two strands of the double-stranded DNA while the 
string of open circles denotes the elongating RNA molecule. The dashed 
lines connecting the circles denote the unbroken non-covalent bonds 
between the complementary subunits on the DNA and RNA strands. Each of 
the grey ovals represents the catalytic site on the corresponding RNAP. 
(b) A simplified version of the fig.\ref{fig-rnaptraf}(a). The DNA track 
for the RNAP motors is assumed to be, effectively, an one-dimensional 
lattice. Each TEC has been replaced by a rectangular black box that 
can cover $r$ lattice sites simultaneously ($r = 6$ in this figure). 
The RNAP in each TEC can exist in either of the two chemical states.
}
\label{fig-rnaptraf}
\end{figure}

For the purpose of quantitative modeling, we simplify the schematic 
picture of RNAP traffic shown in fig.\ref{fig-rnaptraf}(a). We 
represent the DNA track for RNAP motors by a one-dimensional lattice 
and each TEC by a rectangular box (see fig.\ref{fig-rnaptraf}(b)). 
Although the actual size of a TEC may be slightly larger than that 
of the associated RNAP, from now onwards, in this paper we shall 
ignore this size difference. In other words, we assume that the 
size of the black box in the fig.\ref{fig-rnaptraf}(b) is identical 
to that of a TEC as well as that of a RNAP motor. We label the sites 
of the lattice by the integer index $i$ (by convention, from left to 
right). The sites $i=1$ and $i = L$ represent the start and stop sites, 
respectively. Each of the remaining sites in between the start and 
stop sites (i.e., $2 \leq i \leq L-1$) represents a single nucleotide 
on the DNA template. The size of a single RNAP is such that each motor 
can simultaneously cover $r$ successive nucleotides on the DNA 
template (usually, $r$ is typically $30$ to $35$ base pairs, but in 
fig\ref{fig-rnaptraf} $r = 6$). According to our convention, the 
position of each RNAP is denoted by the integer index of the lattice 
site covered by the leftmost site of the RNAP. Thus, the allowed 
range of the positions $j$ of each RNAP is $1 \leq j \leq L$. The 
hard-core steric interactions among the RNAPs is captured by 
imposing the condition that no lattice site is allowed to be covered 
simultaneously by more than one RNAP. Irrespective of the actual 
numerical value of $r$, each RNAP can move forward or backward by only 
one site in each time step, if demanded by its own mechano-chemistry, 
provided the target site is not already covered by any other RNAP. 
This is motivated by the fact that a RNAP must transcribe the 
successive nucleotides one by one. 

The total number of RNAPs on the DNA template is denoted by the symbol 
$N$. Under periodic boundary conditions (PBC), $N$ is independent of 
time whereas $N$ is a fluctuating time-dependent quantity if open 
boundary conditions (OBC) are imposed on the system. Therefore, 
$\rho = N/L$ is the {\it number density} of the RNAPs. The  
{\it coverage density} is defined by $\rho_{cov} = N r/L = \rho~ r$ 
which is the total fraction of the nucleotides covered by all the RNAPs 
together. Under OBC, the number density as well as the coverage density 
are, in general, fluctuating quantities, but the average of these 
densities attain time-independent values in the stationary state.

\begin{figure}[h]
\begin{center}
\includegraphics[angle=-90,width=0.95\columnwidth]{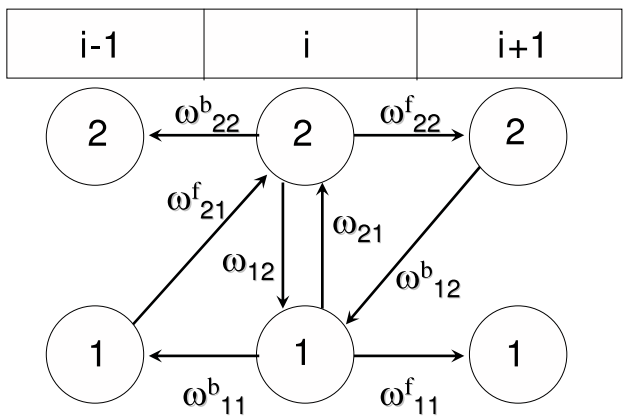}
\end{center}
\caption{ A schematic representation of the mechano-chemical cycle of 
each RNAP in our model in the elongation stage. No $PP_{i}$ is 
bound to the RNAP in the state $1$ whereas the $PP_{i}$-bound state 
of the RNAP is labelled by the index $2$.
}
\label{fig-rnaptm1}
\end{figure}

Our model is aimed at the elongation stage and is not intended to describe 
the initiation and termination processes in detail. Therefore, we represent 
initiation and termination by the two parameters $\alpha$ and $\beta$, 
respectively. Whenever the site $i=1$ on the DNA template is vacant, 
this site is allowed to be occupied by a new RNAP with the probability 
$\alpha$ in the time interval $\Delta t$ (in all our numerical calculations 
we take $\Delta t = 0.001$s). Similarly, a RNAP bound to the site $i = L$ 
is allowed to detach from the template with the probability $\beta$ in the 
time interval $\Delta t$. For convenience, we also define the probabilities 
$\omega_{\alpha}$ and $\omega_{\beta}$ for attachment and detachment, 
respectively. Note that $\omega_{\alpha}$ is related to $\alpha$ by the 
relation $\alpha = 1 - exp(-\omega_{\alpha} \Delta t)$;  $\omega_{\beta}$ 
is related to $\beta$ by a similar relation. 

Following Wang et al.\cite{osterrnap98}, we have a simplified description 
of the chemical (or, conformational) states of each individual RNAP. 
Since release of $PP_{i}$ is the rate limiting step in the process of 
elongation of the mRNA transcript, we consider only two, effectively 
distinct, chemical states of the RNAP in each mechano-chemical cycle 
during the elongation stage. In the  state labelled by the 
integer index $1$ no $PP_{i}$ is bound to the RNAP whereas the 
$PP_{i}$-bound state of the RNAP is labelled by the index $2$. 
The simplified scheme, that captures the essential mechano-chemical 
processes during the mRNA transcript elongation, is shown in 
fig.\ref{fig-rnaptm1} . In this figure,  
$\omega^{f}_{21}$, $\omega^{f}_{11}$ and $\omega^{f}_{22}$ are the rates 
of polymerization of RNA in three diferrent situations, namely, by the 
hydrolysis of nucleotides (i) on the RNAP, (ii) in solution (while no 
$PP_{i}$ is bound to the RNAP and (iii) in solution, while $PP_{i}$ is 
bound to the RNAP. The corresponding rates of reverse transitions, which 
result in depolymerization of the RNA, are denoted by the symbols  
$\omega^{b}_{12}$, $\omega^{b}_{11}$ and $\omega^{b}_{22}$, respectively. 
Finally, $\omega_{21}$ and $\omega_{12}$ are the rates of association and 
dissociation, respectively, of $PP_{i}$. 

``Backtracking'' and ``hypertracking'' of RNAP have been observed in 
{\it in-vitro} single-RNAP experiments \cite{block05,busta07}. 
Effects of backtrackings on transcription has been investigated recently 
by Voliotis et al.\cite{liverpool}. However, the model used by Voliotis 
et al.  \cite{liverpool} does not explicitly capture the biochemical 
transitions of a RNAP during its enzymatic cycle. Interestingly, it has 
been experimentally demonstrated \cite{nudler03a,nudler03b} that 
backtracking of a RNAP gets strongly suppressed if there is another RNAP 
close behind it. Therefore, we do not allow the possibility of 
backtracking in our model as, except at extremely low densities of RNAPs, 
backtrackings and hypertrackings are expected to be rare in RNAP traffic. 

Four different types of nucleotides are used by nature to synthesize 
all DNA molecules. Sequence inhomogeneity can lead to site-dependent 
rates of translocation of RNAP on its track. In the context of TASEP, 
which is a special limit of our model of RNAP traffic, effects of 
quenched random site-dependent hopping rates
\cite{lebowitz,schutzdis,tripathi,goldstein,kolwankar,harris,derrida,santen2,barmarev}, 
have been investigated extensively over the last decade. Moreover, 
Brownian motors with quenched disorder \cite{harms,marchesoni,family,jia} 
have also been studied. In the same spirit, single molecular motors,  
which move on DNA or RNA tracks, have been modelled assuming the 
nucleotide sequence on the track to be random \cite{nelson1,nelson2}.  

However, to our knowledge, for the realistic inhomogeneous, but 
correlated, sequences no analytical technique is available at present 
for the calculation of the quantities of our interest in this paper. 
In fact, the theoretical schemes developed so far for single RNAPs 
\cite{mdwang04,tadigotla}, which take into account the actual sequence 
of the specific DNA track, are implemented numerically. Even in the 
context of earlier models of protein synthesis, almost all the theoretical 
results on the effects of sequence inhomogeneities have been obtained by 
computer simulations \cite{dong,basuchow}. Therefore, for the sake of 
ease of analytical calculations, throughout this paper we have ignored 
the sequence inhomogeneity of the nucleotides on the DNA template and, 
instead, assumed a hypothetical homogeneous sequence. 

Let $P_{\mu}(i,t)$ denote the probability that there is a RNAP at the 
spatial position $i$ and in the chemical state $\mu$ at time $t$; 
$\mu =1$ refers to the state in which the RNAP is not bound to any 
$PP_{i}$ whereas $\mu = 2$ corresponds to the state with bound $PP_{i}$. 
Note that $P(i) = \sum_{\mu=1}^{2} P_{\mu}(i)$ is the probability of 
finding a RNAP at the site $i$, irrespective of its chemical state. 
Similarly, $P_{\mu} = \sum_{i} P_{\mu}(i)$ is the probability of 
finding a RNAP in the chemical state $\mu$ irrespective of its spatial 
position. We describe the stochastic dynamics of the system in terms 
of master equations for $P_{\mu}(i,t)$. Most of our analytical results 
have been derived using the mean-field approximation. 

In order to test the range of validity of our approximate analytical 
calculations, we have also carried out extensive computer simulations 
(Monte Carlo simulations) of our model. In these simulations, we have 
used random sequential updating which appropriately corresponds to the 
Master equations used in our analytical formalisms. In each run of the 
simulations, the system was allowed to reach steady state in the first 
one million time steps and the data for the steady-state were collected 
over the next eight million time steps. The entire process was repated 
with large number of different initial conditions and, finally, 
average steady-state flux was computed. We have observed that 
the qualitative features of our results do not depend significantly on 
the actual numerical value of $r$ as long as is sufficiently larger 
than unity. Therefore, unless stated otherwise, all the numerical 
results plotted in this paper have been obtained taking $r = 10$.
In our test simulation runs, we did not find any significant  
variations in the data for $L \geq 1000$. Therefore, almost all the 
simulation data reported here were generated in our production runs by 
keeping $L$ fixed at $L = 1000$.

\section{RNAP traffic under Periodic Boundary Conditions}
\label{sec-pbc}

We always denote the spatial position of a RNAP on the DNA track by 
the integer index of the site covered by the left edge of the RNAP (i.e., 
the leftmost of the $r$ successive sites representing the RNAP). Thus, 
in our terminology, a site is {\it occupied} by a RNAP if it coincides 
with the leftmost of the $r$ sites representing that RNAP while the next 
$r-1$ sites on its right are said to be {\it covered} by the same RNAP.

Let $P(\underline{i}|j)$ be the conditional probability that, given a 
RNAP at site $i$, there is another RNAP at site $j$; the underlined 
index $i$ within the bracket denotes the site whose occupational status 
is given. Obviously, $Q(\underline{i}|j)$ is the conditional probability 
that, given a RNAP at site $i$, site $j$ is empty; the meaning of the 
underlined index $i$ within the bracket is the same as in case of $P$. 
Note that, if site $i$ is given to be occupied by one RNAP, the site 
$i-1$ can be covered by another RNAP if, and only if, the site $i-r$ is 
also occupied.

\subsection{Mean-field theory under periodic boundary conditions}

In the mean-field approximation, the master equations for $P_{\mu}(i,t)$ 
are given by 
\begin{widetext}
\begin{eqnarray}
\frac{dP_{1}(i,t)}{dt}
&=& ~\omega^{b}_{11} ~P_{1}(i+1,t) ~Q(i+1-r|\underline{i+1}) + ~\omega^{f}_{11} ~P_{1}(i-1,t) ~Q(\underline{i-1}|i-1+r) + ~\omega^{b}_{12} ~P_{2}(i+1,t) ~Q(i+1-r|\underline{i+1}) \nonumber \\
&+& ~\omega_{12} ~P_{2}(i,t) - ~\omega_{21} ~P_{1}(i,t) - ~\omega^{f}_{11} ~P_{1}(i,t) ~Q(\underline{i}|i+r) - ~\omega^{f}_{21} ~P_{1}(i,t) ~Q(\underline{i}|i+r) - ~\omega^{b}_{11} ~P_{1}(i,t) ~Q(i-r|\underline{i})
\label{eq-masterp1}
\end{eqnarray}
\\
\begin{eqnarray}
\frac{dP_{2}(i,t)}{dt} &=& ~\omega^{b}_{22} ~P_{2}(i+1,t) ~Q(i+1-r|\underline{i+1}) + ~\omega^{f}_{22} ~P_{2}(i-1,t) ~Q(\underline{i-1}|i-1+r) + ~\omega^{f}_{21} ~P_{1}(i-1,t) ~Q(\underline{i-1}|i-1+r) \nonumber\\
&+& ~\omega_{21} P_{1}(i,t) - ~\omega_{12} ~P_{2}(i,t) - ~\omega^{f}_{22} ~P_{2}(i,t) ~Q(\underline{i}|i+r) - ~\omega^{b}_{12} ~P_{2}(i,t) ~Q(i-r|\underline{i}) - ~\omega^{b}_{22} ~P_{2}(i,t) ~Q(i-r|\underline{i})
\label{eq-masterp2}
\end{eqnarray}
\end{widetext}
Note that the two equations (\ref{eq-masterp1}) and (\ref{eq-masterp2}) 
are not independent of each other because of the condition
\begin{eqnarray}
P(i) &=& P_{1}(i) + P_{2}(i) = \frac{N}{L} = \rho 
\label{eq-normal1}
\end{eqnarray}

For our numerical calculations, we choose the same set of rate constants 
which Wang et al.\cite{osterrnap98} extracted from empirical data; these 
are as follows:\\
\begin{eqnarray}
\omega^{f}_{21} &=& \omega^{f0}_{21} \cdot [NTP], ~{\rm with} ~\omega^{f0}_{21} = 10^{6} ~M^{-1} \cdot s^{-1} \nonumber \\ 
\omega^{f}_{11} &=& \omega^{f0}_{11} \cdot [NMP], ~{\rm with} ~\omega^{f0}_{11} = 46.6 ~M^{-1} \cdot s^{-1} \nonumber \\
\omega^{f}_{22} &=& \omega^{f0}_{22} \cdot [NMP], ~{\rm with} ~\omega^{f0}_{22} = 0.31 ~M^{-1} \cdot s^{-1} \nonumber \\
\omega_{21} &=& \omega^{0}_{21} \cdot [PP_{i}], ~{\rm with} ~\omega^{0}_{21} = 10^{6} ~M^{-1} \cdot s^{-1} \nonumber \\
\omega_{12} &=& 31.4 ~s^{-1} \nonumber \\
\omega^{b}_{12} &=& 0.21 ~s^{-1} \nonumber \\
\omega^{b}_{11} &=& 9.4 ~s^{-1} \nonumber \\
\omega^{b}_{22} &=& 0.063 ~s^{-1} \nonumber \\
\end{eqnarray}
where we have used the abbreviation NMP for nucleoside monophosphate.

Compared to Basu and Chowdhury's model \cite{basuchow} of ribosome-driven 
protein synthesis, our model of RNAP-driven mRNA synthesis involves 
fewer chemical states and, hence, fewer master equations. In fact, the 
number of chemical states in this model is equal to those in an earlier 
model of traffic of single-headed kinesin motors \cite{nosc,greulich} 
where the two chemical states, however, have totally different physical 
interpretations. But, the number of terms involved in each of the 
master equations \ref{eq-masterp1} and \ref{eq-masterp2} are much larger 
than those in ref.\cite{basuchow} and in ref.\cite{nosc,greulich}.

\subsection{Steady state properties under periodic boundary conditions}

In the steady state all $P_{\mu}(i,t)$  become indepent of time. Moreover, 
because of the PBC, these probabilities are also independent of the site 
index $i$ in the steady state of the system. Therefore, from Bayes's 
theorem, 
\begin{eqnarray}
P(\underline{i}|i+r)&=&\frac{P(i|\underline{i+r}) P(i+r)}{P(i)}\nonumber\\
&=&P(i|\underline{i+r})
\label{eq-condp2}
\end{eqnarray}
and, hence,
\begin{eqnarray}
Q(\underline{i}|i+r) = Q(i|\underline{i+r}). 
\label{eq-condq1}
\end{eqnarray}
We calculate $Q(\underline{i}|i+r)$ along the same line as sketched in 
ref.\cite{basuchow}. Given that the site $i$ is occupied, the conditional 
probability that the site $i+r$ is also occupied is given by
\begin{equation}
  P(\underline{i}|i+r)=\frac{N-1}{L+N-Nr-1}.
\label{eq-condp1}
\end{equation}
Thus, in the limit $L \rightarrow \infty$ and $N \rightarrow \infty$, 
while keeping $\rho = N/L$ fixed, we get 
\begin{eqnarray}
Q(\underline{i}|i+r) = Q(i|\underline{i+r}) = \frac{1-\rho r}{1+\rho-\rho r}
\label{eq-condq2}
\end{eqnarray}
Note that $Q$ vanishes at $\rho = 1/r$, because the entire stretch of the 
DNA template between the points of initiation and termination of 
transcription is fully covered by the RNAPs at $\rho r = \rho_{cov} = 1$.

\begin{figure}[t]
\begin{center}
(a)\\[0.25cm]
\includegraphics[width=0.95\columnwidth]{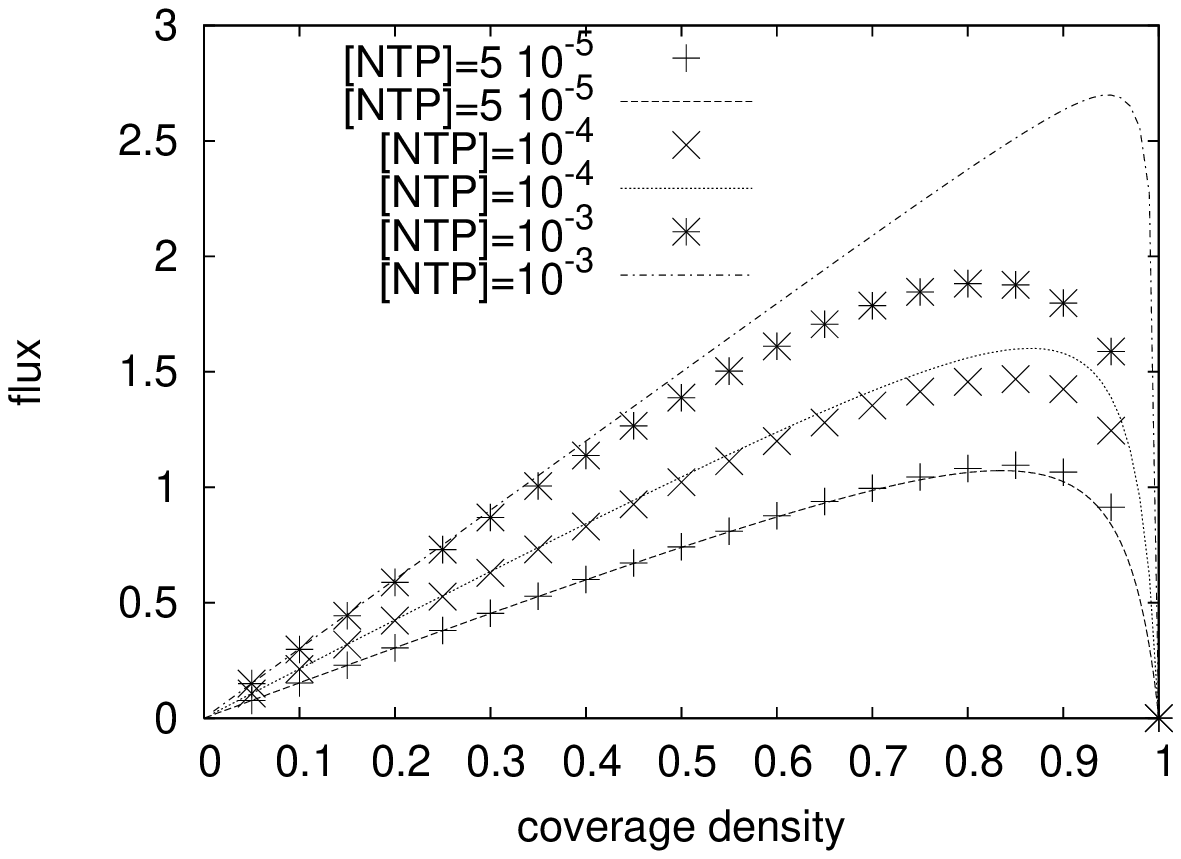}\\[0.5cm]
(b)\\[0.25cm]
\includegraphics[width=0.95\columnwidth]{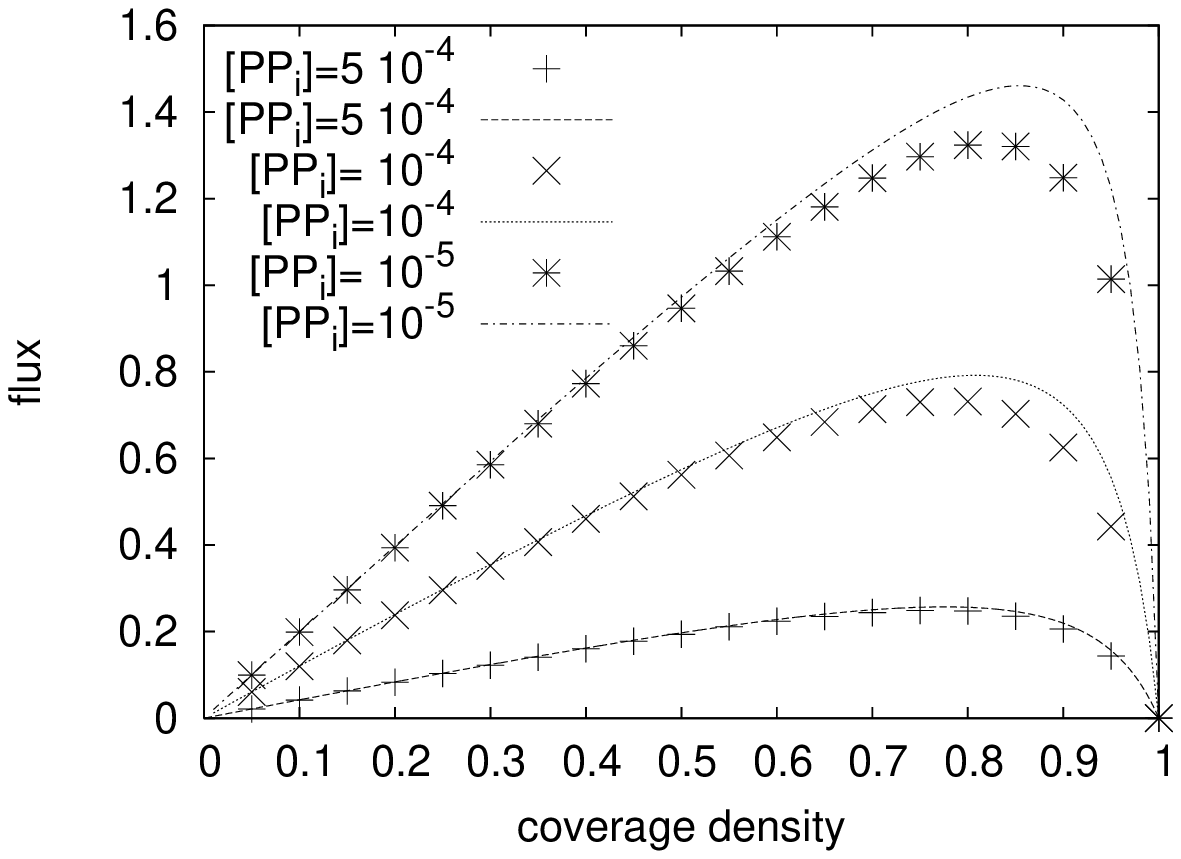}
\end{center}
\caption{The steady-state flux of the RNAPs, under periodic boundary 
conditions, plotted as a function of the coverage density $\rho_{cov}$ for 
(a) three different values of [NTP] at [PP$_i$] = 1 $\mu$M, and 
(b) three different values of [PP$_i$] at [NTP] = 1 mM. 
The lines correspond to our mean-field theoretic predictions 
whereas the discrete data points have been obtained from computer 
simulations. 
}
\label{fig-fdpbc}
\end{figure}

Solving Eqs.(\ref{eq-masterp1}), together with (\ref{eq-normal1}) in the 
steady state under PBC, we get 
\begin{eqnarray}
P_{1} = \biggl(\frac{\omega_{12} + \omega^{b}_{12}Q}{\Omega_{\updownarrow} + \Omega_{\leftrightarrow}Q}\biggr) ~\rho  \nonumber \\
P_{2} = \biggl(\frac{\omega_{21} + \omega^{f}_{21}Q}{\Omega_{\updownarrow} + \Omega_{\leftrightarrow}Q}\biggr) ~\rho
\label{eq-stsolnpbc}
\end{eqnarray}
where 
\begin{eqnarray}
\Omega_{\updownarrow} = \omega_{12} + \omega_{21} \nonumber \\
\Omega_{\leftrightarrow} = \omega^{f}_{21} + \omega^{b}_{12} 
\end{eqnarray}
and $Q$ is given by the equation (\ref{eq-condq2}).

In the steady state under PBC, the flux of the RNAPs is given by 
\begin{eqnarray}
J &=& ~(\omega^{f}_{11} + \omega^{f}_{21}) ~P_{1} ~Q(\underline{i}|i+r) + ~\omega^{f}_{22} ~P_{2} ~Q(\underline{i}|i+r)\nonumber\\
&-& ~\omega^{b}_{11} ~P_{1} ~Q(i-r|\underline{i}) - ~(\omega^{b}_{22} + \omega^{b}_{12}) ~P_{2} ~Q(i-r|\underline{i}) \nonumber \\
\label{eq-fluxdef}
\end{eqnarray}
Hence,
\begin{eqnarray}
J &=& ~\Omega_{1} ~P_{1} ~Q + ~\Omega_{2} ~P_{2} ~Q \nonumber \\
&=& (~\Omega_{1} ~P_{1} + ~\Omega_{2} ~P_{2}) \biggl(\frac{1-\rho_{cov}}{1+\rho-\rho_{cov}}\biggr) 
\label{eq-stfluxpbc}
\end{eqnarray}
where 
\begin{eqnarray} 
\Omega_{1} = \omega^{f}_{11} + \omega^{f}_{21} - \omega^{b}_{11} \\
\Omega_{2} = \omega^{f}_{22} - \omega^{b}_{12} - \omega^{b}_{22}, 
\end{eqnarray}
are two {\it effective forward hopping rates} from the states $1$ and $2$, 
respectively, while $Q$ is given by the equation (\ref{eq-condq2}).
Since $P_{1} = 0 = P_{2}$ at $\rho=0$ the corresponding flux $J$ vanishes. 
$J$ also vanishes at $\rho_{cov} = 1$ as $Q$ vanishes at $\rho r = 1$. 

Our mean-field estimate (\ref{eq-stfluxpbc}) of flux $J$ is plotted 
against the coverage density $\rho_{cov}$ in fig.\ref{fig-fdpbc} for 
(a) three different values of $[NTP]$ at $[PP_i] = 1 \mu M$, and 
(b) three different values of $[PP_i]$ at $[NTP] = 1 mM$. 
The qualitative features of these fundamental diagrams are similar to 
those observed earlier \cite{basuchow} in the context of ribosomal 
traffic during protein synthesis from a mRNA template. The most notable 
feature of these diagrams is their asymmetric shape. This shape 
of the fundamental diagram is in sharp contrast to the symmetry of the 
fundamental diagram of TASEP about $\rho = 1/2$. The physical reason 
for the asymmetric shape of the fundamental diagram in fig.\ref{fig-fdpbc}  
is the same as in ribosomal traffic \cite{basuchow}. 

The rate constant $\omega^{f}_{21}$ is higher at higher concentrations 
of NTP and gives rise to higher flux, i.e., higher rate of transcriptional 
output (see fig.\ref{fig-fdpbc}). Conversely, higher concentration of 
$PP_{i}$ opposes the release of $PP_{i}$ thereby slowing down the overall 
rate of transcription. Moreover, at higher concentrations of NTP each RNAP 
attempts forward stepping more frequently; while making these attempts, 
it feels stronger hindrance at higher densities of RNAPs. Therefore, 
the deviation of the mean-field estimates of flux from the corresponding 
simulation data is larger at higher NTP concentration and at higher 
coverage density of the RNAPs. Similarly, forward stepping of a RNAP 
is less suppressed when the $PP_i$ concentration in the solution is 
lower; therefore, stronger devitation of the mean-field estimates of flux 
from the corresponding simulation data is observed at lower $PP_i$ 
conentration and higher RNAP densities.

\section{Results under Open boundary Conditions}
\label{sec-obc}

Open boundary conditions are more realistic than PBC for describing RNAP 
traffic during transcription. A fresh RNAP can attach with the site $i=1$ 
only in the state $1$ (i.e., no $PPi$ is bound to it).
In this section we make a further assumption for simplifying the equations. 
We replace the conditional probability $Q(\underline{i}|j)$, by the 
probability $Q(j)$ that site $j$ is empty, irrespective of the state of 
occupation of any other site. Note that the probability of finding a 
``hole'' at $j$ (i.e., the probability that the site $j$ is not ``covered'' 
by any RNAP) is given by $1-\sum_{s=0}^{r-1}P(j-s)$.

\subsection{Mean-field theory under open boundary conditions}

Under mean-field approximation, the master equations for the probabilities 
are  now given by 
\begin{widetext}
\begin{eqnarray}
\frac{dP_{1}(1,t)}{dt} &=& ~\omega_{\alpha} ~\biggl(1-\sum_{s=1}^{r}~P(s)\biggr)
+ ~\omega^{b}_{11} ~P_{1}(2,t) + ~\omega^{b}_{12} ~P_{2}(2,t) + ~\omega_{12} ~P_{2}(1,t) \nonumber \\
&-& ~ \omega_{21} ~P_{1}(1,t) - (~\omega^{f}_{11} + ~\omega^{f}_{21}) ~P_{1}(1,t) 
\biggl(\frac{1-\sum_{s=1}^{r} P(1+s)}{1-\sum_{s=1}^{r} P(1+s) + P(1+r)}\biggr)  
\label{eq-obc1}
\end{eqnarray}
\begin{eqnarray}
\frac{dP_{1}(i,t)}{dt}
&=& \biggl[~\omega^{b}_{11} ~P_{1}(i+1,t) + ~\omega^{b}_{12} ~P_{2}(i+1,t)\biggr]  \biggl(\frac{1-\sum_{s=1}^{r} P(i+1-s)}{1-\sum_{s=1}^{r} P(i+1-s) + P(i+1-r)}\biggr) \nonumber \\
&+& ~\omega^{f}_{11} ~P_{1}(i-1,t) \biggl(\frac{1-\sum_{s=1}^{r} P(i-1+s)}{1-\sum_{s=1}^{r} P(i-1+s) + P(i-1+r)}\biggr) + ~\omega_{12} ~P_{2}(i,t) \nonumber \\
&-& ~\omega_{21} ~P_{1}(i,t) - (~\omega^{f}_{11} + ~\omega^{f}_{21}) ~P_{1}(i,t) \biggl(\frac{1-\sum_{s=1}^{r} P(i+s)}{1-\sum_{s=1}^{r} P(i+s) + P(i+r)}\biggr)  \nonumber \\
&-& ~\omega^{b}_{11} ~P_{1}(i,t) \biggl(\frac{1-\sum_{s=1}^{r} P(i-s)}{1-\sum_{s=1}^{r} P(i-s) + P(i-r)}\biggr) ~~(i\neq L,i\neq 1)  
\label{eq-obc2}
\end{eqnarray}
\begin{eqnarray}
\frac{dP_{1}(L,t)}{dt} &=&  ~\omega^{f}_{11} ~P_{1}(L-1,t) \biggl(\frac{1-\sum_{s=1}^{r} P(L-1+s)}{1-\sum_{s=1}^{r} P(L-1+s) + P(L-1+r)}\biggr)  
+ ~\omega_{12} ~P_{2}(L,t) \nonumber \\
&-& ~\omega_{21} ~P_{1}(L,t) - ~\omega^{b}_{11} ~P_{1}(L,t) \biggl(\frac{1-\sum_{s=1}^{r} P(L-s)}{1-\sum_{s=1}^{r} P(L-s) + P(L-r)}\biggr) 
- ~\omega_{\beta} ~P_{1}(L,t) 
\label{eq-obc3}
\end{eqnarray}
\begin{eqnarray}
\frac{dP_{2}(i,t)}{dt} &=& \biggl[~\omega^{f}_{21} ~P_{1}(i-1,t) + ~\omega^{f}_{22} ~P_{2}(i-1,t)\biggr] \biggl(\frac{1-\sum_{s=1}^{r} P(i-1+s)}{1-\sum_{s=1}^{r} P(i-1+s)+P(i-1+r)}\biggr) \nonumber\\
&+& ~\omega^{b}_{22} ~P_{2}(i+1,t) \biggl(\frac{1-\sum_{s=1}^{r} P(i+1-s)}{1-\sum_{s=1}^{r} P(i+1-s) + P(i+1-r)}\biggr) \nonumber \\
&+& ~\omega_{21} ~P_{1}(i,t)  - ~\omega_{12} ~P_{2}(i,t) 
- ~\omega^{f}_{22} ~P_{2}(i,t) \biggl(\frac{1-\sum_{s=1}^{r} P(i+s)}{1-\sum_{s=1}^{r} P(i+s) + P(i+r)}\biggr) \nonumber\\
&-& (~\omega^{b}_{12} + ~\omega^{b}_{22}) ~P_{2}(i,t) \biggl(\frac{1-\sum_{s=1}^{r} P(i-s)}{1-\sum_{s=1}^{r} P(i-s) + P(i-r)}\biggr) ~~~~~~~~~~~(i\neq L)
\label{eq-obc4}
\end{eqnarray}
\begin{eqnarray}
\frac{dP_{2}(L,t)}{dt} &=& \biggl[~\omega^{f}_{21} ~P_{1}(L-1,t) + ~\omega^{f}_{22} ~P_{2}(L-1,t)\biggr]   \biggl(\frac{1-\sum_{s=1}^{r} P(L-1+s)}{1-\sum_{s=1}^{r} P(L-1+s) + P(L-1+r)}\biggr) \nonumber\\
&+& ~\omega_{21} ~P_{1}(L,t)  - ~\omega_{12} ~P_{2}(L,t) - (~\omega^{b}_{12} + ~\omega^{b}_{22}) ~P_{2}(L,t) \biggl(\frac{1-\sum_{s=1}^{r} P(L-s)}{1-\sum_{s=1}^{r} P(L-s) + P(L-r)}\biggr) \nonumber \\
&-& ~\omega_{\beta} ~P_{2}(L,t)
\label{eq-obc5}
\end{eqnarray}
\end{widetext}

\subsection{Steady state properties under open boundary conditions}

\begin{figure}[t]
\begin{center}
(a)\\[0.25cm]
\includegraphics[width=0.95\columnwidth]{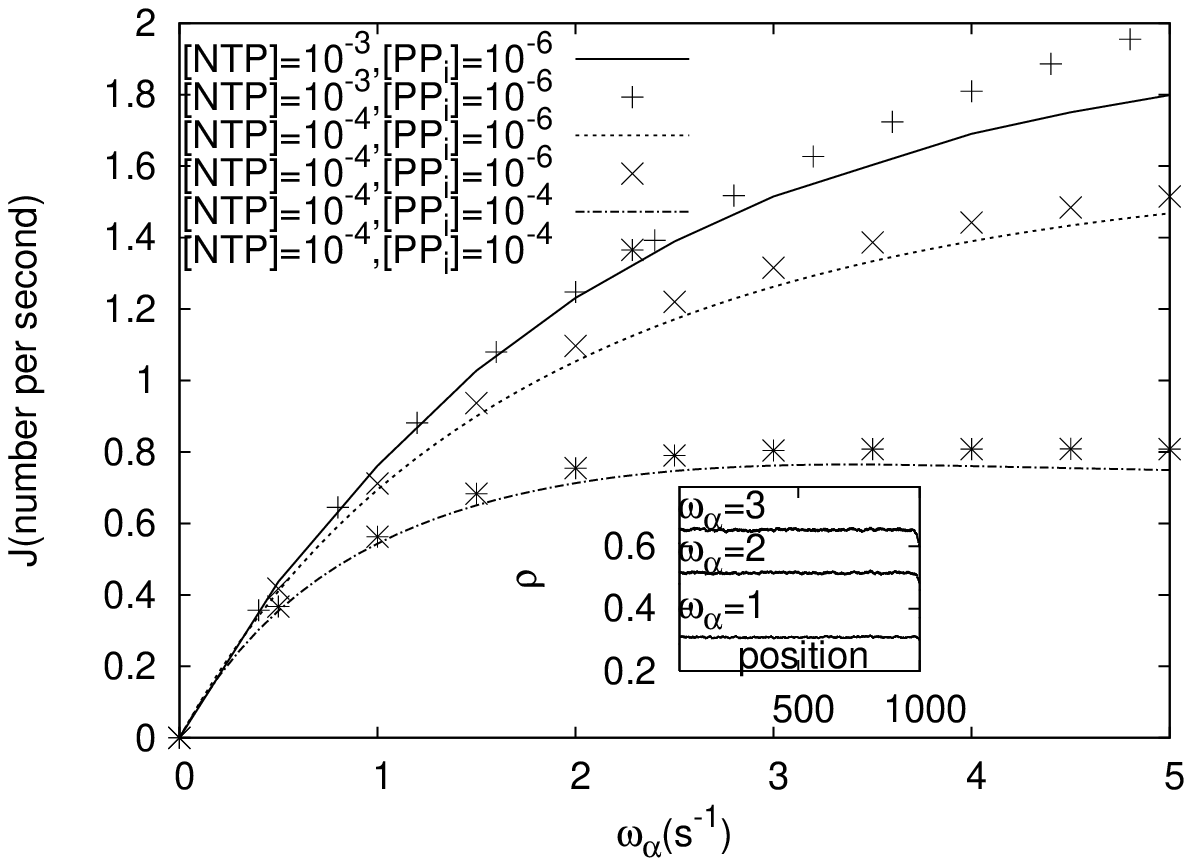}\\[0.5cm]
(b)\\[0.25cm]
\includegraphics[width=0.95\columnwidth]{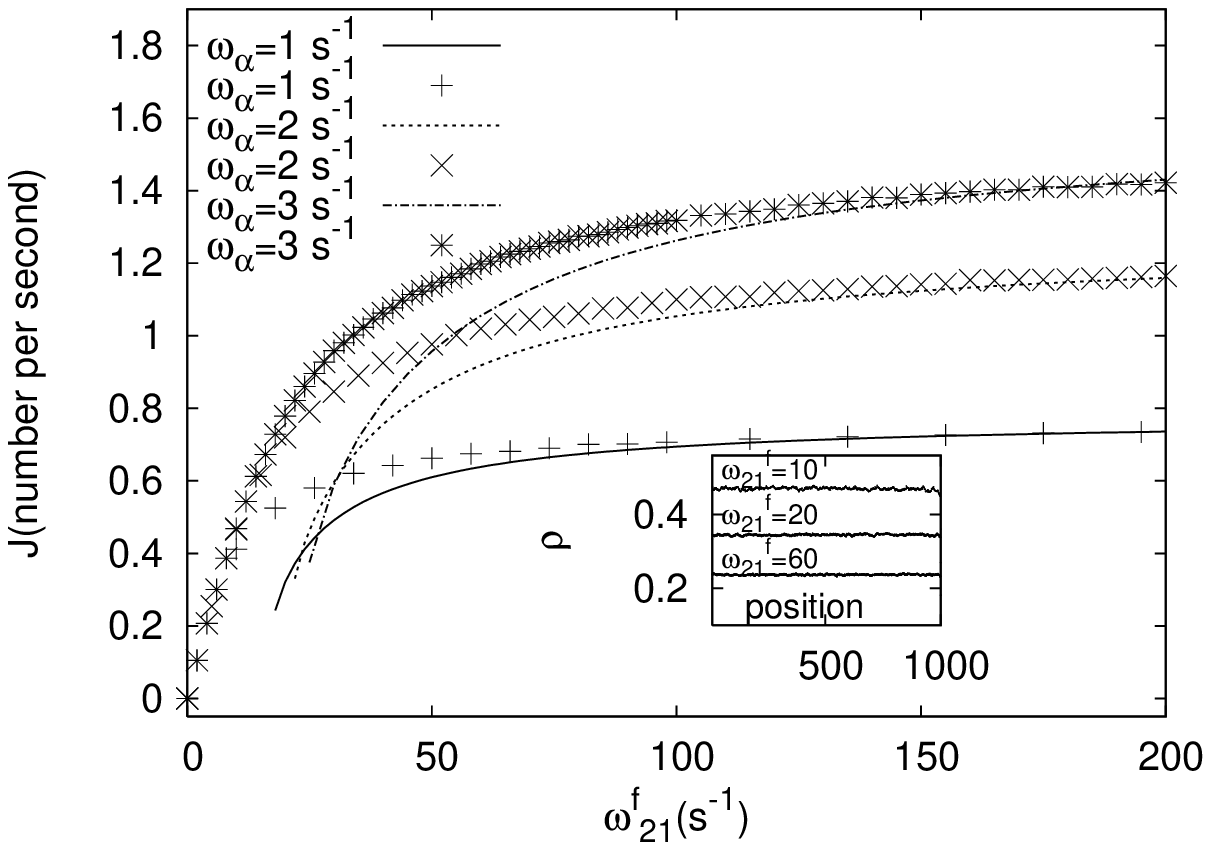}
\end{center}
\caption{The steady-state flux of the RNAPs, under open boundary 
conditions, plotted as a function of  
(a) $\omega_{\alpha}$,  for three sets of values of the pair of 
parameters  [NTP], [PP$_i$]; 
(b) $\omega^{f}_{21}$ for three values of the parameter $\omega_{\alpha}$. 
The lines correspond to our mean-field theoretic predictions 
whereas the discrete data points have been obtained from computer 
simulations. The insets show the average density profiles for 
(a) three different values of $\omega_{\alpha}$ and (b) three different 
values of $\omega^{f}_{21}$.
}
\label{fig-fdobc}
\end{figure}

Using these mean-field equations (\ref{eq-obc1}-\ref{eq-obc5}) in the 
steady state, we have numerically calculated our theoretical estimates 
of the flux. These mean-field theoretic estimates are plotted as 
functions of the rate constants $\omega_{\alpha}$ and $\omega^{f}_{21}$, 
respectively, in figs.\ref{fig-fdobc}(a) and (b). In order to test the 
level of accuracy of these approximate theoretical predictions, we have 
compared these results with the corresponding numerical data obatined 
from our direct computer simulations of the model under open boundary 
conditions. We have also computed the average density profiles and 
plotted these profiles for three different values of $\omega_{\alpha}$ 
and three different values of $\omega^{f}_{21}$ in the insets of 
figs.\ref{fig-fdobc}(a) and (b), respectively. 

The flux increases monotonically with increasing $\omega_{\alpha}$ 
as well as with increasing $\omega^{f}_{21}$ and, eventually, saturates  
in both the cases. This trend of variation of flux is accompanied by a 
monotonic {\it rise} of the average density profile of the RNAPs in 
fig.\ref{fig-fdobc}(a) and with a monotonic {\it fall} of the average 
density profile in fig.\ref{fig-fdobc}(b). A comparison of these  
qualitative features of the variation of flux and density profiles 
with those in ribosome traffic \cite{basuchow}, indicates a transition 
from the low-density phase to the maximal current phase in 
fig.\ref{fig-fdobc}(a) and from the high-density phase to the maximal 
current phase in fig.\ref{fig-fdobc}(b) \cite{basuchow,schuetz}.

\begin{figure}[t]
\begin{center} 
(a)\\[0.25cm]
\includegraphics[width=0.95\columnwidth]{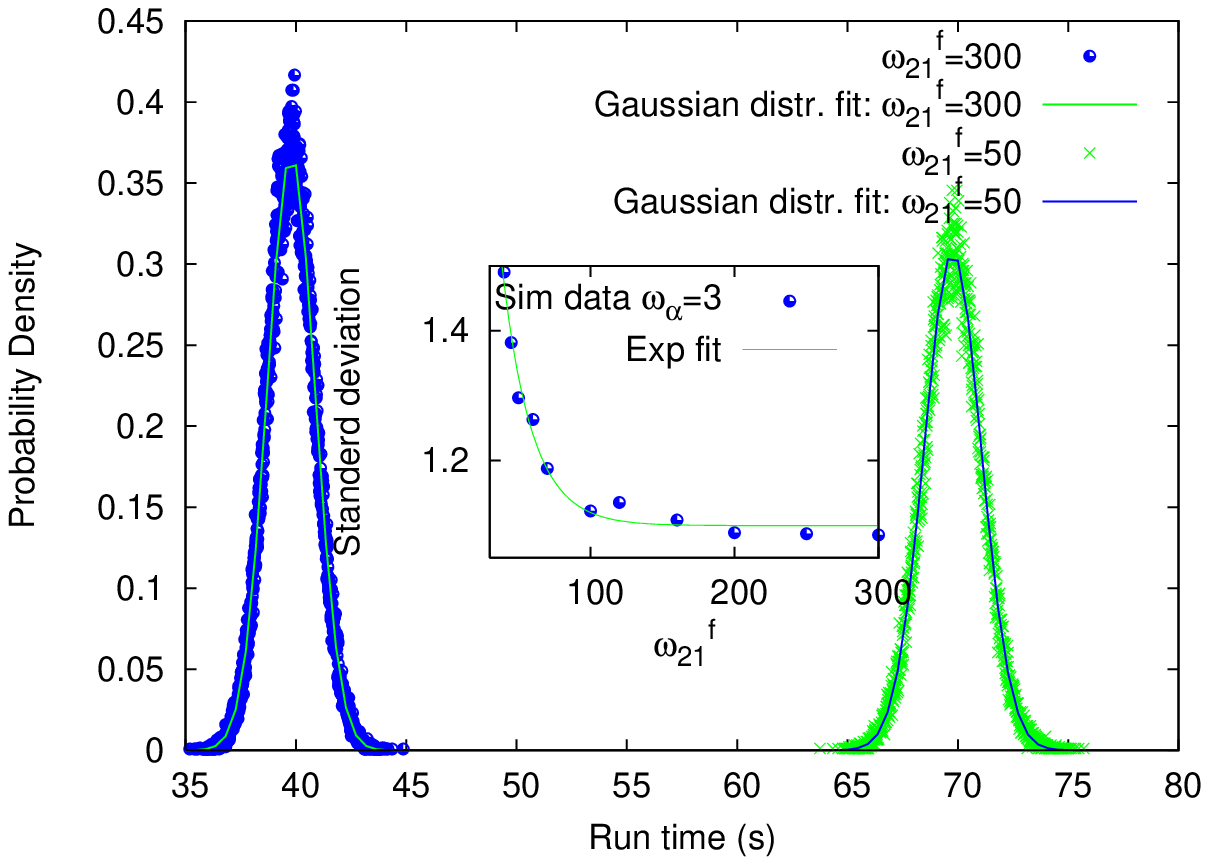}\\[0.5cm]
(b)\\[0.25cm]
\includegraphics[width=0.95\columnwidth]{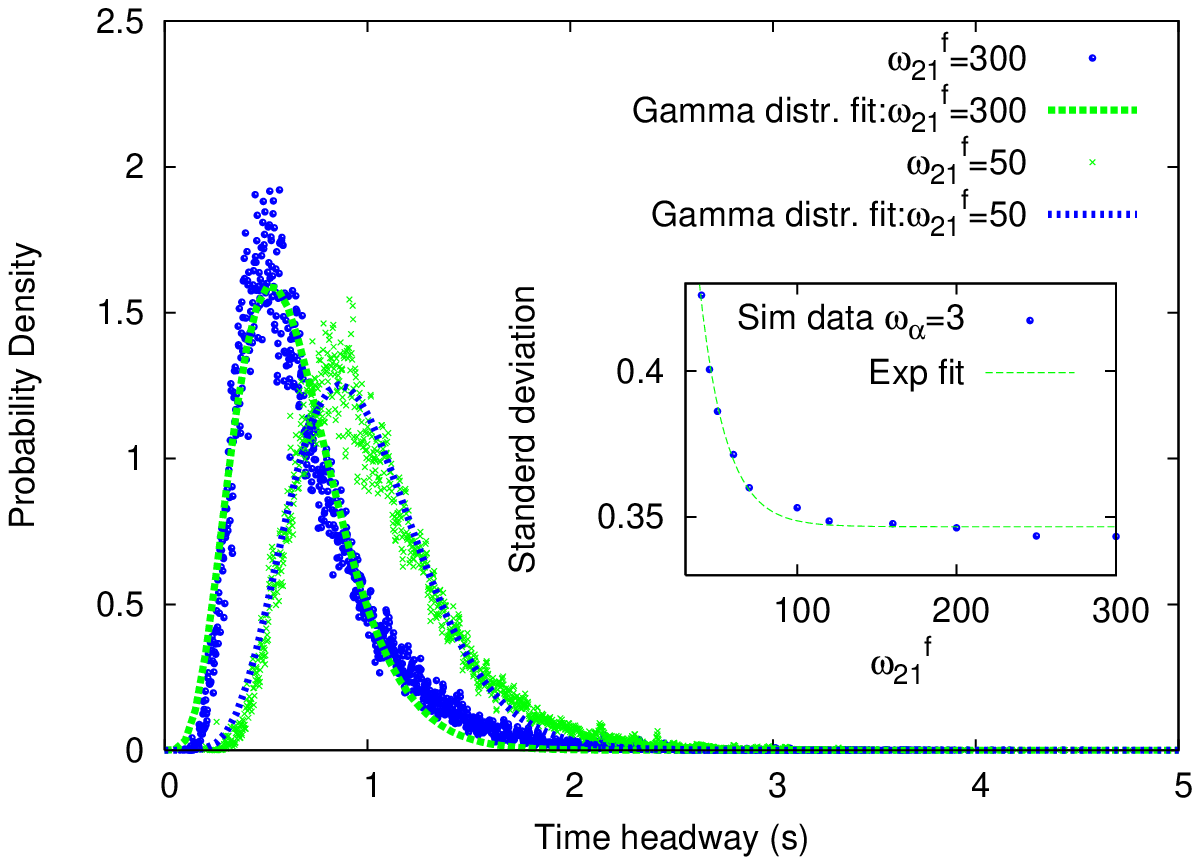}
\end{center}
\caption{The distributions of the run times and time headways in our 
model, under open boundary conditions, plotted in (a) and (b), 
respectively, for two different values of $\omega^{f}_{21}$. 
The discrete data points have been obtained from computer simulations. 
The curves fitted to these data points are drawn with the lines. 
The variation of the standard deviations of the distributions of 
run times and time-headways with the increase of the parameter 
$\omega^{f}_{21}$ are shown in the insets; the discrete points have 
been obtained from the simulation data and the best fit curve through 
these points has been drawn by a line.
}
\label{fig-fluc}
\end{figure}

\begin{figure}[t]
\begin{center} 
(a)\\[0.25cm]
\includegraphics[width=0.95\columnwidth]{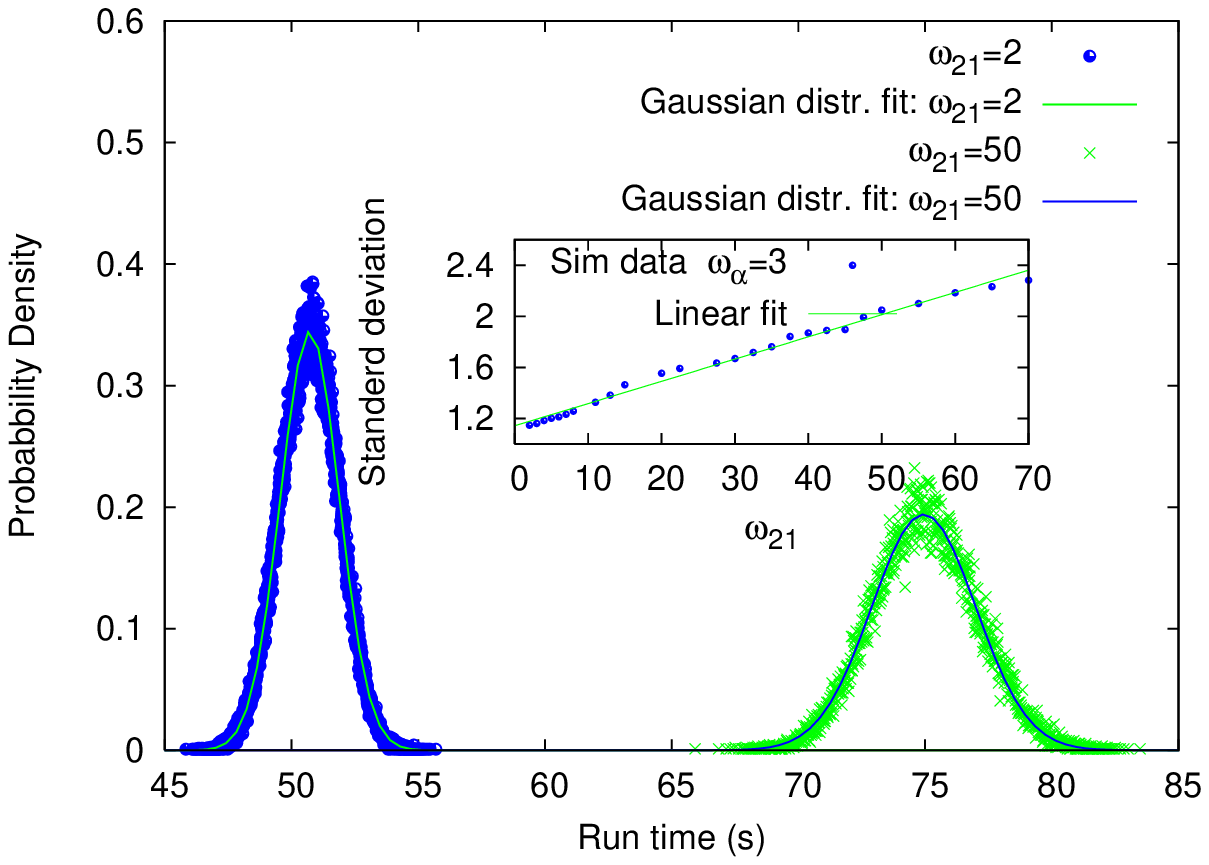}\\[0.5cm] 
(b)\\[0.25cm]
\includegraphics[width=0.95\columnwidth]{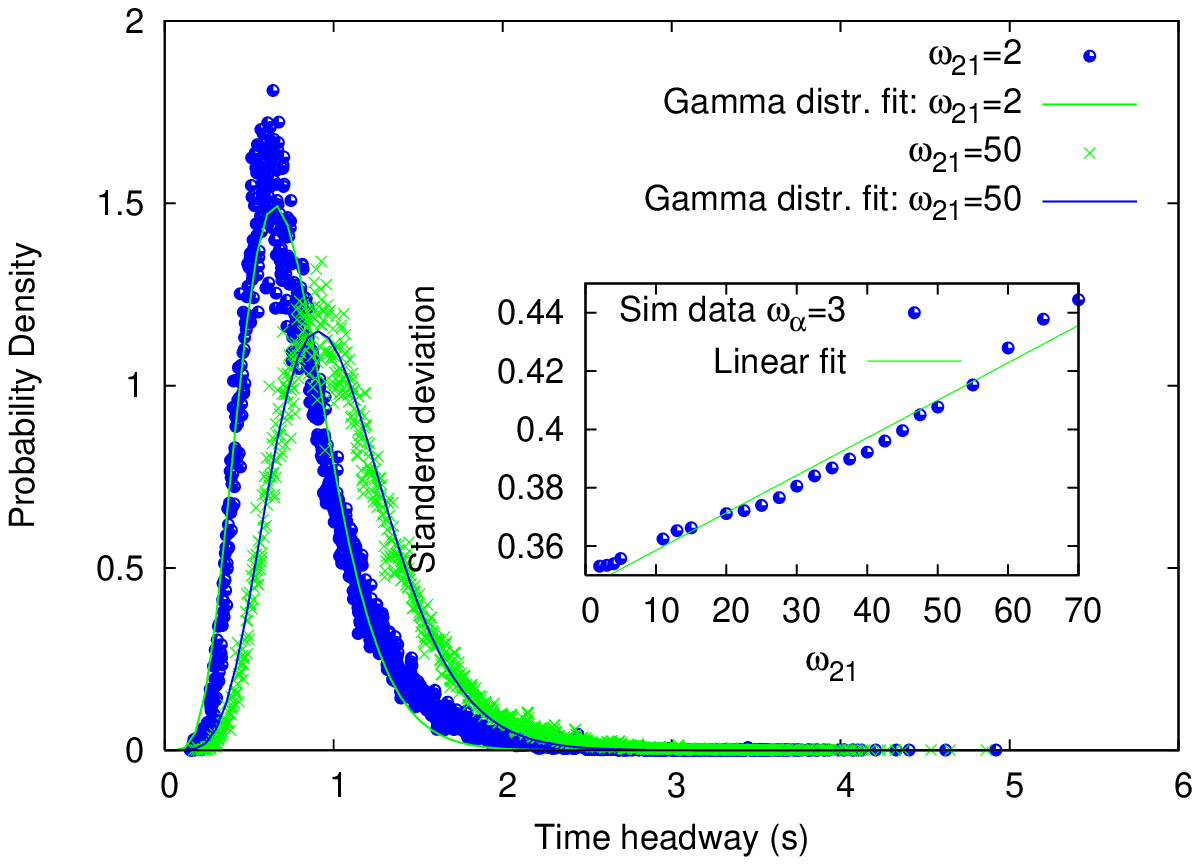}
\end{center}
\caption{The distributions of the run times and time-headways in our 
model, under open boundary conditions, plotted in (a) and (b), 
respectively, for two different values of $\omega_{21}$. 
The discrete data points have been obtained from computer simulations. 
The curves fitted to these data points are drawn with the lines. 
The variation of the standard deviation of the distributions of 
run times with the increase of the parameter $\omega_{21}$ are shown 
in the insets; the discrete points have been obtained from the 
simulation data and the best fit curve through these points has been 
drawn by a line.
}
\label{fig-noise}
\end{figure}

\section{RNAP-to-RNAP fluctuations and transcriptional noise}
\label{sec-fluc}

The distribution $\tilde{P}_{T}$ of run times $T$ is a measure of 
the RNAP-to-RNAP fluctuations in the rates of transcription. This 
distribution, obtained from computer simulations of our model, is 
plotted in fig.\ref{fig-fluc}(a) for two different values of the 
parameter $\omega^{f}_{21}$. The gaussian fit to the distribution 
$\tilde{P}_{T}$ is consistent with the gaussian distributions of 
the ``delay times'' obtained by Morelli and J\"ulicher \cite{morelli07} 
in the limit of sufficiently large number of intermediate steps. 
Gaussian distributions of the speeds of the RNAPs were observed 
by Tolic-Norrelykke et al.\cite{tolic} in their {\it in-vitro} 
experiments. Although this conclusion in ref.\cite{tolic} was based 
on the assumption of uniform speed of the RNAPs during the elongation 
stage, what was actually observed in their experiments is the Gaussian 
distribution of the run times; this is certainly consistent with 
our theorertical result.  

We define the standard deviation, i.e., the root-mean-square deviations 
\begin{equation}
\eta_{T} = <(T - <T>)^2>^{1/2} 
\end{equation}
of run times $T$ from their mean, as a measure of the transcriptional 
noise arising from the stochastic mechano-chemical cycles of the RNAPs. 
$\eta_{T}$ is plotted as a function of $\omega^{f}_{21}$ in the inset 
of fig.\ref{fig-fluc}(a). Since 
$\omega^{f}_{21} = \omega^{f0}_{21} \dot [NTP]$, 
the inset of fig.\ref{fig-fluc}(a) clearly establishes that the 
transcriptional noise $\eta_{T}$ falls exponentially with increasing 
concentration of NTP. This trend of variation is consistent with the 
well known fact that the fluctuations in the rates of chemical 
synthesis are stronger when the concentrations of reactants are lower.

The distribution ${\cal P}_{\tau}$ of time-headways $\tau$ is a measure 
of the fluctuations in the time interval between the completion of the 
polymerization of successive mRNA transcripts. This distribution, also 
obtained from computer simulations of our model, is plotted in 
fig.\ref{fig-fluc}(b) for the same values of the $\omega^{f}_{21}$ as 
those used in fig.\ref{fig-fluc}(a). The best fit to the numerical data 
for ${\cal P}_{\tau}$ is of the general form
\begin{equation} 
{\cal P}_{\tau} = C \tau^{\nu} e^{-\mu \tau}
\label{eq-gamatype}
\end{equation}
with positive constants $\mu$ and $\nu$; $C$ being the normalization 
constant. The form (\ref{eq-gamatype}) is consistent with the gamma 
distribution that is expected for the time-headways at suffciently 
low densities. 

We define 
\begin{equation}
\eta_{\tau} = <(\tau - <\tau>)^2>^{1/2}
\end{equation}
as a measure of the fluctuations in the time-headways. In the inset of 
fig.\ref{fig-fluc}(b) we plot $\eta_{\tau}$ as a function of 
$\omega^{f}_{21}$; the best fit to this curve is an exponential.

In the limit in which $\omega_{12} \rightarrow \infty$ and all other 
rate constants, except $\omega^{f}_{21} = q$, vanish, our model reduces 
to TASEP if, simultaneously $r \rightarrow 1$. In this limit of our 
model ${\cal P}_{\tau}$ is expected to be well approximated by the 
exact expression for TASEP with {\it parallel} updating \cite{thchow1,thchow2}: 
\begin{widetext}
\begin{eqnarray}
{\cal P}_{\tau} =  
\left[\frac{qy}{\rho-y}\right] \{1-(qy/\rho)\}^{t-1} +
 \left[\frac{qy}{(1-\rho)-y}\right] \{1-(qy/(1-\rho))\}^{t-1} 
 - \left[\frac{qy}{\rho-y}+\frac{qy}{(1-\rho)-y}\right] p^{t-1} - q^2(t-1)p^{t-2}. \nonumber \\
\end{eqnarray} 
\label{eq-thtasep}
\end{widetext}
where 
\begin{equation}
y = \frac{1}{2q}\left( 1 - \sqrt{1 - 4 q \rho (1-\rho)}\right).
\end{equation}

Finally, the transcriptional noise increases, instead of decreasing, 
with the increase of $PP_{i}$ concentration (see fig.\ref{fig-noise}); 
in other words, increase of $PP_{i}$ concentration not only slows 
down the average rate of RNA synthesis, but also makes transcription 
more noisy.

\section{Implications for experiments}
\label{sec-expt}

Almost all the quantitative theories of RNAP developed so far  
\cite{julicherrnap98,osterrnap98,sousa96,sousa97,sousa06,mdwang04,mdwang07,nudler05,tadigotla,peskinrnap06,woo06} 
were intended to account for the mechano-chemistry of a {\it single} RNAP. 
The interactions of RNAPs in transcriptional interference \cite{shearwin05}
is a well known phenomenon and it has also been modelled quantitatively
\cite{sneppen05}. However, instead of studying interactions of RNAPs 
during the transcription of different genes 
\cite{nudler03a,nudler03b,crampton06,sneppen05}, we have modelled the 
steric interactions of RNAPs which are simultaneously involved in the 
transcription of the same gene.

The possibility of steric interactions of RNAPs during their traffic-like 
collective movements along the same DNA template has been known for a 
long time \cite{alberts,jackson98}. The ``christmas-tree''-like structures 
\cite{scheer97,raska04} observed in electron microscopic studies of 
eukaryotic transcription arise from simultaneous transcription of the same 
gene by many RNAPs. These structures also have strong similarity with 
the dense population of the nascent mRNA transcripts observed all along 
the loops of the DNA strands in the electron micrographs of lampbrush 
chromosomes \cite{callan86,gall06}. 

Our theory predicts not only the average rate of synthesis of RNA, but 
also two different measures of fluctuations in the process of 
transcription. In most of the earlier experimental investigations of 
transcriptional noise, the distributions of the sizes and frequencies 
of the ``burst'' of the transcriptional activity were recorded. However, 
size and frequency of the bursts depend on the temporal resolution used 
for {\it sorting} the time series of the events into separate bursts 
(see, for example, Fig.1 of ref.\cite{cox06}). Therefore, in principle, 
the statistics of reported distributions of burst sizes and frequencies 
may change with the change of time resolution selected for such sorting. 
Instead, in this paper, we have introduced new measures of the 
stochasticity in transcriptional activity which do not require any 
sorting of this kind. 

In recent years sophisticated optical techniques have been developed 
for single mRNA imaging \cite{tal04,rodriguez07,darzacq07b,goldingpre}. 
We believe that our theoretical predictions can be tested most 
appropriately by carrying out {\it in-vitro} experiments with either 
fluorescently labelled RNAPs \cite{tolic} or using techniques for tagging 
the nascent mRNA with fluorescent probes \cite{tal04,golding05} or using 
techniques where fluorescent probes can quickly bind with the nascent 
mRNA as soon as it is released by the RNAP \cite{raj}. Comparison of 
our theoretical predictions on the distributions of run times and 
time headways require collection of appropriate data. In our theory, the 
run time includes the time spent by a RNAP in the elongation stage as well 
as in the termination stage, but does not include the time spent in the 
initiation stage. Therefore, in experiments, run times of the RNAps 
should be measured only from the instant when the TEC gets stabilized; a 
technique used in ref.\cite{tolic} may be utilized for this purpose.

\section{Summary and conclusions}
\label{sec-conclude}

Surprisingly, no attempt has been made in the past to develope mathematical 
models for RNAP traffic where transcription of a single gene is carried out 
simultaneously by a stream of RNAPs closely spaced on the same DNA template.  
To our knowledge, the model developed in this paper is the first attempt 
to capture inter-RNAP interactions in a model where the mechano-chemical 
cycles of each individual RNAPs in the elongation stage are also 
incorporated, albeit in a simplified manner. In analogy with vehicular 
traffic \cite{css}, we have defined the flux for RNAP traffic; the 
RNAP flux is also the total rate of synthesis of RNA. We have calculated 
the average rates of RNA synthesis analytically under mean-field 
approximation. 

Drawing analogies with vehicular traffic, we have defined two novel 
quantities whose distributions serve as measures of RNAP-to-RNAP 
fluctuations in the transcription of a single gene. We have calculated 
these distributions numerically by carrying out computer simulations 
of our model. The widths of these distributions (more precisely, 
root-mean-square fluctuations) can be treated as good measures of the 
strength of ``transcriptional noise''. We have investigated how the 
level of ``transcriptional noise'' depends on some of the model 
parameters which can be varied in a controlled manner in laboratory 
experiments. A similar analysis of ``translational noise'', which 
arises from ribosome-to-ribosome fluctuations during protein synthesis 
from the same mRNA template, will be reported elsewhere \cite{garaietal}.
The inhomogeneous sequence of nucleotides on the DNA template can lead 
to stronger fluctuations thereby making additional contributions to the 
levels of transcriptional noise. The ``intrinsic noise'' studied in 
this paper arises from the stochastic nature of the steps of the 
mechano-chemical cycle of individual RNAPs. Although the noise level gets 
affected by the interactions of the RNAPs, this noise remains relevant 
even when the gene is transcribed by one RNAP at a time. We have made 
concrete suggestions as to the experimental systems and techniques 
which, in principle, can be used to test our theoretical predictions.\\


\noindent{\bf Acknowledgements}: One of the authors (DC) thanks Ido 
Golding, Stephan Grill, Anatoly Kolomeisky, Tannie Liverpool, Alex 
Mogilner, Evgeny Nudler, George Oster, Arjun Raj and Gunter Sch\"utz 
for useful comments and/or suggestions on the manuscript. DC also 
thanks Bruce Alberts, Frank J\"ulicher and Arjun Raj for drawing his 
attention to some earlier experimental works. DC also thanks Martin 
Depken, Eric Galburt, Debasis Kundu, Luis Morelli, T.V. Ramakrishnan, 
Beate Schmittmann and Simon F. Tolic-Norrelykke for enlightening 
discussions. DC acknowledges hospitality of the Research Center 
J\"ulich and the Max-Planck Institute for the Physics of Complex 
Systems (MPI-PKS Dresden) where parts of this manuscript were prepared. 
This work is supported (through DC) by a research grant from CSIR 
(India) and the Visitors Program of MPI-PKS Dresden (Germany).


\end{document}